\documentclass{article}

\usepackage{graphics} 
\usepackage{subfigure}

\newcommand{\mib}[1]{\mbox{\boldmath$#1$}}

\begin{document}

\title{Reaction cross sections of proton scattering from carbon isotopes (A=8-22)
by means of the relativistic impulse approximation
}

\author{Kaori Kaki \\
\it Department of Physics, Shizuoka University, Shizuoka 422-8529, Japan \\
and
\\
\it RIKEN Nishina Center, Wako 351-0198, Japan}
\maketitle

\begin{abstract}%
Reaction cross sections of carbon isotopes for proton scattering are calculated in large energy
region.
Density distributions of carbon isotopes are obtained from relativistic mean field results.
Calculations are based on the relativistic impulse approximation,
and results are compared with experimental data.
A strong relationship between reaction cross section and root-mean-square radius is clearly shown 
for $^{12}$C using a model distribution.
\end{abstract}


\section{Introduction}
Unstable nuclei are fruitful objects for nuclear physics
because they provide new information about halo structure, magic numbers, 
nuclear matter properties and many other things which are very different from stable nuclei.
In order to obtain such information on the unstable nuclei,
proton-elastic scattering is expected to be the most appropriate experiment
besides electron scattering.
The experimental observables are, however, restricted  due to their unstableness,
and reaction cross section or interaction cross section is considered at first.
For unstable nuclei: $^{6,8}$He, $^{11}$Li, and $^{11,14}$Be, 
the interaction cross sections have been found to be significantly large, 
and have provided their large root-mean-square radii \cite{tani85, tani92}.
The reaction cross section have been calculated for proton-rich isotopes, e.g.
carbon \cite{hor07,ibr08}, helium, lithium, beryllium, oxygen, and nitrogen in addition to 
carbon \cite{shar13} in terms of the Glauber model based on the Glauber theory \cite{gl59} and/or the eikonal model.
Refs.\cite{hor07, shar13} have considered unstable nuclei scattering from $^{12}$C
target, and Ref.\cite{ibr08} has calculated proton-nucleus scattering.
Nuclear structures for the unstable nuclei have been provided by
different ways:
a Slater determinant generated from a phenomenological mean-field potential \cite{hor07,ibr08} , 
and relativistic\cite{pat91} and nonrelativistic mean-filed theories in Ref.\cite{shar13}.
Reaction cross sections for $^{14,15,16}$C isotopes scattering from $^{12}$C target
have been studied based on the g-matrix double-folding model and $^{14}\rm C+n$
two-body model \cite{tm14}.
In addition to the neutron-rich isotopes,  the angular distribution of proton-elastic scattering from
the proton-rich isotopes like $^9$C has been measured \cite{ym13}.

In the present study, densities of carbon isotopes are provided by  
the relativistic mean-field (RMF) results \cite{hs81,st94}, and reaction cross sections
for proton-elastic scattering from carbon isotopes are calculated
with two prescriptions: the Glaubar model and the relativistic impulse approximation
(RIA) \cite{tw87a,tw87b}.
As for target carbon nucleus, the mass numbers A=8 through A=22 are considered, e.i.,
proton-rich isotopes are also included besides neutron-rich ones.
The former isotopes are expected to have a large root-mean-square radius
due to the Coulomb repulsive interaction between protons.
It has been well known that the large reaction cross section or interaction cross section
provides the large root-mean-square radius for the neutron-rich nuclei,
however, for the proton-rich nuclei, the relationship between such radius and the reaction cross section is expected to be different from the relation with respect to the neutron-rich
isotopes since the number of proton itself does not increase, and the NN amplitudes of
proton-proton scattering are different frmo proton-neutron scattering.
The present study, therefore, pays attention to the relationship between 
the root-mean-square
radius and the reaction cross section for $^{8-22}$C nuclei.

There is another interest in obtaining the information on the structure
of unstable nuclei besides the root-mean-square radius
form the restricted observables.
The author has proposed a procedure for calcium \cite{kh99,kk09}, 
nickel \cite{kk15} isotopes, and  $^{208}$Pb \cite{kk04},
in which the density distribution is
assumed by the Woods-Saxon function, and two parameters of the function
are determined with two observables.
In the present study, showing a specific relationship between the reaction cross section 
and the root-mean-square radius is attempted for carbon isotopes
rather than determining the density distributions.

In Sec. II, formulas of RIA on which the analysis is based are briefly presented.
Numerical results are given in Sec. III,
In addition to the reaction cross sections, 
results calculated with RIA for proton-elastic scattering from $^{12,13,14}$C,
and comparison between numerical results and experimental data
are also given in Sect. III.
The summary and conclusion of present study appear in Sec. IV.

\section{Formulation}

The Dirac equation containing the optical potential is described in momentum space as follows:
\begin{eqnarray}
&& \{ \gamma^0 E - \mib{\gamma} \cdot \mib{p}' - m \} \ \Psi(\mib{p}')
- \int \frac{d^3p}{(2 \pi)^3} \ \hat U(\mib{p}',\mib{p}) \ \Psi(\mib{p}) = 0,
\nonumber \\
&&
\label{eq:a}
\end{eqnarray}
where $\Psi(\mib{p})$ is given by the Fourier transformation of the wave function in coordinate
space:
\begin{eqnarray}
&& \Psi(\mib{p}) = \int d^3r e^{-i \mib{p} \cdot \mib{r}} \Psi(\mib{r}),
\nonumber \\ 
&& 
\label{eq:b}
\end{eqnarray}
where natural unit ( $\hbar = c = 1$ ) is taken.

In accordance with the prescription of the RIA \cite{tw87a,tw87b}, 
the Dirac optical potential is given in momentum space by
\begin{eqnarray}
\hat{U}(\mib{p}',\mib{p}) & = & - \frac{1}{4} {\rm Tr}_2 
\left\{ \int \frac{d^3k}{(2 \pi)^2} 
\hat{M}_{pp}(\mib{p},\mib{k}-\frac{\mib{q}}{2} \rightarrow \mib{p}',\mib{k}+\frac{\mib{q}}{2} )
\ \hat{\rho}_p(\mib{k},\mib{q}) \right\} \nonumber \\ 
                          &   & - \frac{1}{4} {\rm Tr}_2 
\left\{ \int \frac{d^3k}{(2 \pi)^2} 
\hat{M}_{opt}(\mib{p},\mib{k}-\frac{\mib{q}}{2} \rightarrow \mib{p}',\mib{k}+\frac{\mib{q}}{2} )
\ \hat{\rho}_n(\mib{k},\mib{q}) \right\},
\end{eqnarray}
where 
$\hat{\rho}_p$ and $\hat{\rho}_n$
are density matrices for protons and neutrons, respectively.
The trace is over the $\gamma$ matrices with respect to the target nucleons and
the subscript 2 in the trace corresponds to the target nucleons.

As discussed in Ref.\cite{tw87b} it is known that the nuclear density generally
varies more rapidly with $\mib k$ than the NN amplitude and is largest at $\mib k=0$.
Therefore taking the optimal factorization into account, the optical potentials are written in
the well-known $t \rho$ forms:
\begin{eqnarray}
\hat{U}(\mib{p}',\mib{p}) & = & - \frac{1}{4} {\rm Tr}_2 
\left\{ 
\hat{M}_{pp}(\mib{p},-\frac{\mib{q}}{2} \rightarrow \mib{p}',\frac{\mib{q}}{2} )
\ \hat{\rho}_p(\mib{q}) \right\} \nonumber \\ 
                          &   & - \frac{1}{4} {\rm Tr}_2 
\left\{  
\hat{M}_{opt}(\mib{p},-\frac{\mib{q}}{2} \rightarrow \mib{p}',\frac{\mib{q}}{2} )
\ \hat{\rho}_n(\mib{q}) \right\}.
\end{eqnarray}
The relativistic density matrix $\hat \rho$ 
depends only on the momentum transfer $q$, as follows:
\begin{eqnarray}
\hat \rho( \mib{q} )  =  \rho_S(q)+ \gamma_2^0 \rho_V(q) - \frac{i \mib{\alpha}_2 \cdot \mib{q}}{2m} \rho_T(q),                     
\end{eqnarray}
where each term is a Fourier transformation of a coordinate-space density; 
\begin{eqnarray}
\rho_S(q) & = & 4 \pi \int_0^{\infty} \ j_0(qr) \rho_S(r) r^2 dr, \\
\rho_V(q) & = & 4 \pi \int_0^{\infty} \ j_0(qr) \rho_V(r) r^2 dr, \\
\rho_T(q) & = &-4 \pi m \int_0^{\infty} \ \frac{j_1(qr)}{q} \rho_T(r) r^2 dr. 
\end{eqnarray}

Nuclear densities, provided by the relativistic mean-field theory \cite{st94}, 
are described in terms of upper and lower components as follows:
\begin{eqnarray}
\rho_S(r) & = & \sum_{\alpha} \frac{2j+1}{4 \pi} \left[ G_{\alpha}^2(r)-F_{\alpha}^2(r) \right], \\
\rho_V(r) & = & \sum_{\alpha} \frac{2j+1}{4 \pi} \left[ G_{\alpha}^2(r)+F_{\alpha}^2(r) \right],  
\label{eq.vd}
\\
\rho_T(r) & = & \sum_{\alpha} \frac{2j+1}{4 \pi} \left[ 4 G_{\alpha}(r) \times F_{\alpha}(r) \right], 
\end{eqnarray}
where $\alpha$ represents the quantum numbers of the target nucleus.

In the generalized RIA\cite{tw87a,tw87b} the Feynman amplitude for $NN$ scattering is expanded in terms of 
covariant projection operators $\Lambda^{\rho}(p)$ to separate positive ($\rho=+1$) and negative ($\rho=-1$)
energy sectors of the Dirac space.
The invariant amplitudes, $M_n^{\rho_1 \rho_2 \rho_1'\rho_2'}$, and kinetic covariants,
$\kappa_n$, are given by
\begin{eqnarray}
\hat M (p_1,p_2 \rightarrow p_1',p_2' ) =
      \sum_{\rho_1, \rho_2, \rho_1', \rho_2'}
      \Lambda^{\rho_1'}(p_1') \Lambda^{\rho_2'}(p_2') 
\times
      \sum_{n=1}^{13} M_n^{\rho_1 \rho_2 \rho_1' \rho_2'} \
      \kappa_n \Lambda^{\rho_1}(p_1) \Lambda^{\rho_2}(p_2),
\label{eq:c}
\end{eqnarray}
where
subscripts 1 and 2 correspond to the projectile and target nucleons, respectively.
The covariant projection operator $\Lambda^{\rho}(p)$ is defined by 
$\Lambda^{\rho}(p) = \frac{1}{2m}( \rho \ \gamma^{\mu}p_{\mu}+m)$, and
kinetic covariants $\kappa_n$ are constructed from the Dirac matrices.
The scalar Feynman amplitude, $M_n^{\rho_1 \rho_2 \rho_1'\rho_2'}$, consists of the direct
and exchange parts, each of which represents a sum of four Yukawa
terms characterized by coupling constants, masses and cutoff masses.
In the present calculation, the IA2 parametrization of Ref. \cite{tw87a,tw87b} is used.

By substituting Eq.(\ref{eq:b}) into Eq.(\ref{eq:a}) and replacing the momenta with appropriate operators,
the coordinate-space Dirac equation is obtained as
\begin{eqnarray}
\left\{ \gamma^0 E + i \mib{\gamma} \cdot \nabla -m -\tilde U(\mib{r}) \right\} \tilde \Psi(\mib{r}) = 0,
\label{eq:d} \\
\nonumber
\end{eqnarray}
where $\tilde U(\mib{r})$ has five potential terms as in Ref.\cite{tw87b} and is described as follows,
\begin{eqnarray}
\tilde U(\mib{r}) &=& \tilde S(r) + \gamma^0 \tilde V(r) 
                    - i \mib{\alpha} \cdot \hat {\mib{r}} \tilde T(r) \nonumber \\
                  &-& \{ \tilde S_{LS}(r) + \gamma^0 \tilde V_{LS}(r) \} 
                      \ \mib{\sigma} \cdot \mib{L}.
\end{eqnarray}
The local form of the optical potential is obtained by the prescriptions given in Ref. \cite{tw87b},
namely the asymptotic value of the momentum operator and the angular averaged expression 
for nucleon exchange amplitudes, 
which have been expected to be rather good at high energy scattering.

Equation (\ref{eq:d}) is written as two coupled equations for the upper  ( $\tilde \psi_U$ )
and lower  ( $\tilde \psi_L$ ) components, and solving for $\tilde \psi_U$ and using the form
$\tilde \psi(\mib r)_U = K(r) \phi(\mib r)$ in order to remove the first derivative terms
yields the following Schr\"odinger equation for $\phi(\mib r)$:
\begin{eqnarray}
&& \left\{ -\nabla^2 + 2E( U_{ce} + U_{ls} \ \mib \sigma \cdot \mib L) \right\}
\phi(\mib r ) \nonumber \\
&& \hspace{2cm} 
= \left\{ (E-V_C)^2 -m^2 \right\} \phi(\mib r), 
\end{eqnarray}
where Coulomb potential $V_C$ is explicitly written.
Although the IA2 potentials are used, it may be useful to display the form of the potentials
for the simpler IA1 case.
The Schr\"odinger equivalent potentials for IA1 parametrization are given as follows:
\begin{eqnarray}
U_{ce} & = & \frac{1}{2E} \left\{ 2EV + 2mS - V^2 + S^2 -2VV_C \nonumber \right. \\
    &&  + \left. \left( T^2 -\frac{T}{A} \frac{\partial A}{\partial r}+2 \frac{T}{r}
             + \frac{\partial T}{\partial r} \right) \right. \nonumber \\
    && \left. + \left( - \frac{1}{2r^2A} \frac{\partial}{\partial r} 
                     \left( r^2 \frac{\partial A}{\partial r} \right)
             + \frac{3}{4A^2} \left( \frac{\partial A}{\partial r} \right)^2 \right) \right\}, 
\nonumber \\
&& \\
U_{ls} & = & \frac{1}{2E} \left\{ - \frac{1}{rA} \left( \frac{\partial A}{\partial r} \right)
                                  + 2 \frac{T}{r} \right\}, \\
A & = & \frac{1}{E+m} \left( E-V+m+S-V_C \right).
\end{eqnarray}
This IA1 parametrization corresponds to well-known five-term expansion 
and is obtained by setting $\rho_i = \rho_i'=+1 \ (i=1,2)$ and $n_{max}=5$ 
in Eq.(\ref{eq:c}), instead of $n_{max}=13$.
In this case $K(r)=\sqrt{A}$, and comes to 1 as $r \rightarrow \infty$.

\section{Results}
\subsection{Root-mean-square radii and density distributions}

Figure 1 provides
proton and neutron distributions for $^{8-22}$C except $^{21}$C
as it has been already known that $^{21}$C is unbound.
Distributions are vector densities calculated with Eq.(\ref{eq.vd}) and correspond to
hadron densities in the relativistic expression.
Results are provided by relativistic mean field calculations \cite{st94}.
In Fig.1, the panels of (a) and (b) are for $^{8-14}$C and $^{15-22}$C, respectively, 
and solid lines show neutron distributions and dashed lines proton ones.
It  is seen that neutron density is expanding with increasing mass number while proton density almost stays for $^{14-22}$C, and becomes expanding for $^{8-11}$C.

\begin{figure*}
\begin{center}
\begin{tabular}[b]{c}
\subfigure[Results for $^{8-14}$C.]
{\scalebox{0.70}{\includegraphics{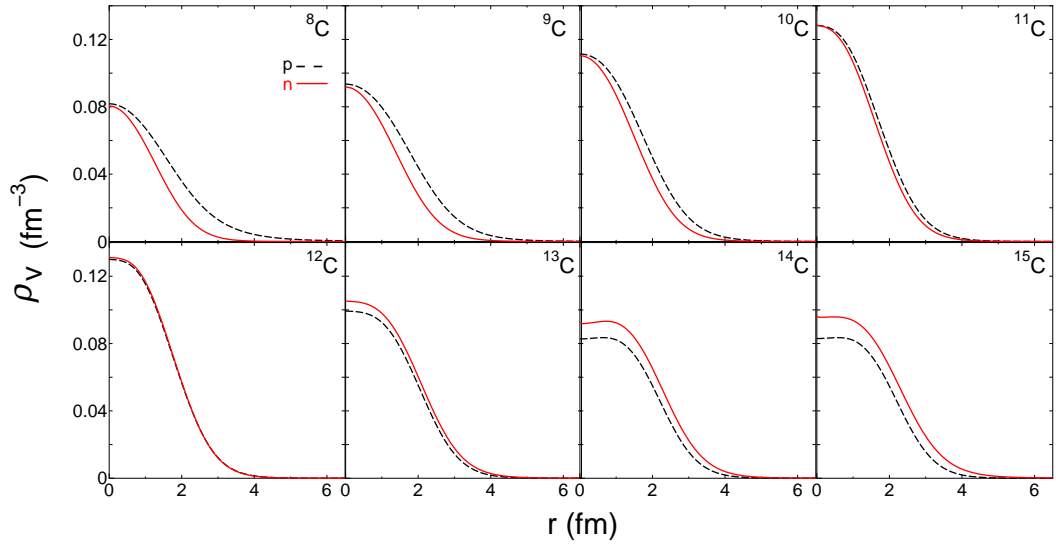}}} \\
\subfigure[Results for $^{15-22}$C.]
{\scalebox{0.70}{\includegraphics{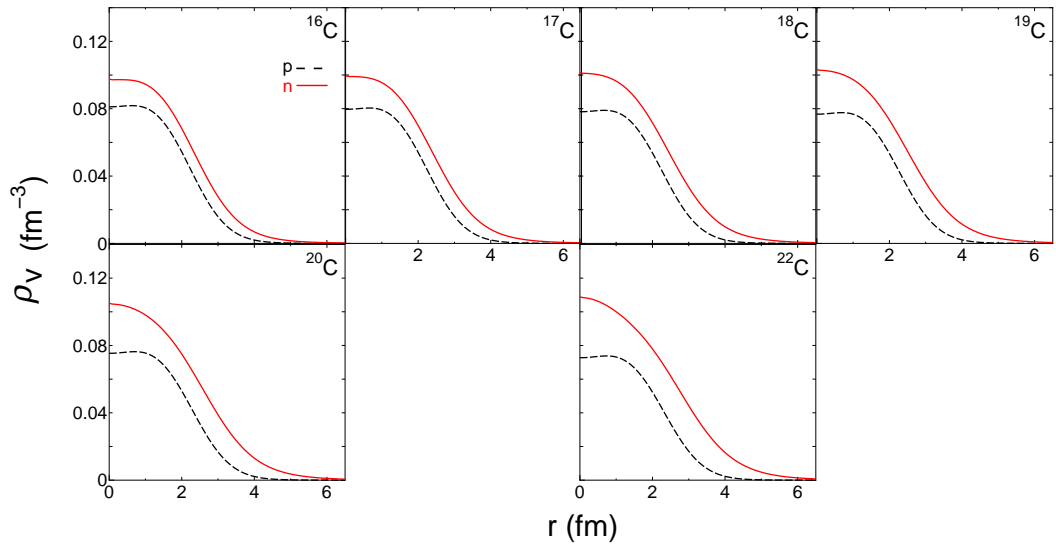}}}

\end{tabular}
\end{center}
\caption{ Proton and neutron distributions.
Solid lines are results for neutron and ashed lines are the results for proton.}            
\end{figure*}

Table 1 shows root-mean-square radius of  proton, neutron,
matter, or charge distribution for C isotopes, respectively.
Values are also provided by relativistic mean field calculations \cite{st94}.
As compared to the average values given in Ref. \cite{am13}, which
are shown in the sixth column, 
the charge radius in Tab.1 is about 2 \% smaller for $^{12}$C, 
and about 3 \% larger for $^{13,14}$C.
The root-mean-square radius of proton is almost flat for neutron-rich carbon isotopes, 
however slightly increasing with increasing mass number. 
In this mass number region, the radius of neutron increases reasonably with increasing mass number.
On the other hand, for the proton-rich carbon isotopes, the root-mean-square radius of proton is increasing drastically with decreasing mass number, and the radius of neutron is almost flat for $^{9-11}$C, and is rather small for $^8$C.
This is reasonable behavior concerning the Coulomb interaction among protons,
the number of which is much larger than that of neutrons.
As a result, the root-mean-square radius of matter density becomes quite large though the mass number is small.

\begin{table}
\caption{Root-mean-square radius of proton ,neutron , matter or charge distribution
for $^{8-22}$C. Three values in the sixth column are charge radii taken from Ref.\cite{am13}.
}
\begin{center}
\begin{tabular}{cccccc} \hline 
         & \multicolumn{4}{c}{Root-mean-square radius (fm)} \\ \cline{2-6}
Isotope  & proton  & neutron  & matter  & charge  & Ref. \cite{am13} \\ \hline
$^8$C &     3.662 & 1.990 & 3.324 & 3.751 \\
$^9$C &     2.872 & 2.189 & 2.664 & 2.984 \\
$^{10}$C & 2.558 & 2.218 & 2.428 & 2.684 \\
$^{11}$C & 2.363 & 2.214 & 2.296 & 2.498 \\
$^{12}$C & 2.277 & 2.257 & 2.267 & 2.417 & 2.4702 \\
$^{13}$C & 2.398 & 2.533 & 2.472 & 2.532 & 2.4614 \\
$^{14}$C & 2.440 & 2.652 & 2.563 & 2.571 & 2.5025 \\
$^{15}$C & 2.445 & 2.944 & 2.755 & 2.576 \\
$^{16}$C & 2.453 & 3.109 & 2.881 & 2.584 \\
$^{17}$C & 2.461 & 3.231 & 2.982 & 2.592 \\
$^{18}$C & 2.470 & 3.323 & 3.065 & 2.600 \\
$^{19}$C & 2.479 & 3.394 & 3.134 & 2.608 \\
$^{20}$C & 2.488 & 3.451 & 3.193 & 2.617 \\
$^{22}$C & 2.507 & 3.539 & 3.289 & 2.635 \\  \hline
\end{tabular}
\end{center}
\label{t1}
\end{table}

In Fig.2 the root-mean-square radii of Tab.1 are shown with respect to mass
number of carbon isotopes.
Solid circles, squares, and triangles are results for proton, neutron, and matter
densities, respectively. Open ones are corresponding to values appeared in Ref. \cite{ibr08} for $^{12-22}$C.
It is found that shell effects of neutrons are significantly small in results for relativistic mean-field calculations, especially in larger mass number.

\begin{figure}
\begin{center}
{\scalebox{0.65}{\includegraphics{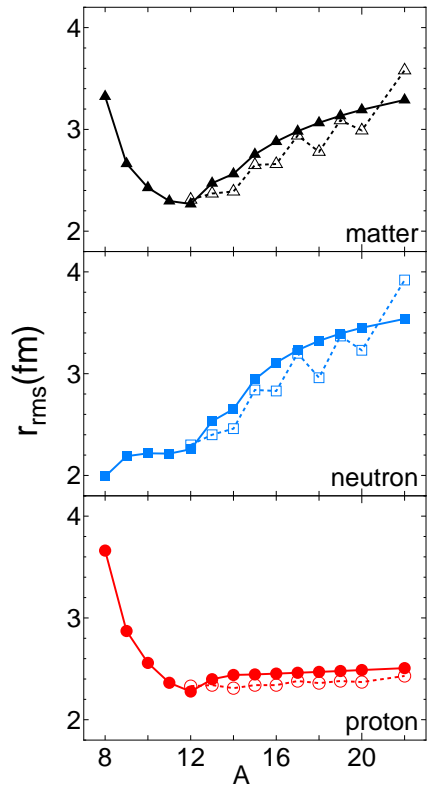}}}
\caption{ Root-mean-square radius with respect to mass number of carbon isotopes.Solid circles, squares, and triangles show results for proton, neutron, and matter
densities, respectively. Open ones are corresponding to values 
which appear in Ref. \cite{ibr08}.}            
\end{center}
\end{figure}

\subsection{Reaction corss sections}
Calculated values of reaction cross section are given in Table 2.
The lowest energy in Tab.II is chosen as same as in the case of Ref.\cite{kk14},
where reasonable results have been obtained in such low energy for He isotopes, 
while  the RIA calculations are available for the proton incident energies more
than 50 MeV.

\begin{table}
\caption{Reaction cross sections calculated based on RIA for $^{8-22}$C.
The unit is square fm.}
\begin{center}
\begin{tabular}{ccccccccc} \hline 
         & \multicolumn{8}{c}{Proton incident energy  (MeV)} \\ \cline{2-9}
Isotope  & 71   & 100  & 200  &300  & 425 & 550 & 650 &800 \\ \hline
$^8$C   &24.18 &24.73&18.87&17.88&18.99&21.33&23.47&24.52 \\
$^9$C   &26.47&26.65&20.49&19.18&19.80&21.68&23.28&24.01 \\
$^{10}$C &28.45&28.35&21.97&20.38&20.62&22.17&23.40&23.91 \\
$^{11}$C &30.17&29.85&23.30&21.47&21.40&22.68&23.62&23.95 \\
$^{12}$C &32.21&31.75&24.84&22.78&22.47&23.61&24.38&24.61 \\
$^{13}$C &36.32&35.86&27.64&25.27&24.81&26.11&27.03&27.35 \\
$^{14}$C &39.32&38.79&29.79&27.17&26.55&27.88&28.82&29.15 \\
$^{15}$C &43.26&42.70&32.43&29.46& 28.62&30.01&31.00&31.35 \\
$^{16}$C &46.71&46.11&34.81&31.55&30.51&31.94&32.97&33.33 \\
$^{17}$C &49.98&49.33&37.11&33.58&32.35&33.82&34.88&35.25 \\
$^{18}$C &53.06&52.36&39.32&35.52&34.13&35.63&36.71&37.10 \\
$^{19}$C &55.98&55.21&41.44&37.40&35.85&37.38&38.48&38.88 \\
$^{20}$C &58.74&57.90&43.47&39.22&37.52&39.07&40.19&40.60 \\
$^{22}$C &63.87&62.89&47.33&42.67&40.70&42.30&43.45&43.87 \\  \hline
\end{tabular}
\end{center}
\label{t2}
\end{table}

In order to compare the RIA results for $^{8-11}$C with the Glauber results,
the reaction cross sections for all carbon isotopes considered here are calculated with 
vector densities obtained from RMF results in accordance with the procedure
of Ref.\cite{ibr08}., and are referred as the Glauber calculations.

Figure 3  shows reaction cross sections as a function of the mass number
at energies: 100, 425, and 800 MeV, respectively.
Solid circles are results for RIA, ,and solid triangles for the Glauber calculation.
Open triangles are corresponding to the results for Ref.\cite{ibr08}, in which
nuclear density distributions are merely different from the present Graluber calculation.
As expected, mass number dependence of the reaction cross section between
solid and open triangles is similar to that is seen in the root-mean-square
radius of matter or neutron distribution in Fig.2.
In Fig.3, the RIA calculation always gives larger values than Glauber calculation, 
and such difference between them seems to come from the difference of the NN interactions based on the calculations.
In both calculations, the reaction cross sections for neutron rich isotopes reasonably increase with increasing mass number or the root-mean-square radius of the matter density distributions to which neutron densities mainly contribute.
And the reaction cross sections for proton rich isotopes do not show large value corresponding to the large matter radius.
One of the reasons why such thing occur is that the cross section of pp-scattering 
is smaller than that of pn-scattering in low energy region, and comes to almost similar in high energy region.
Therefore expanding proton distribution which gives large root-mean-square radius, dose not contribute to the reaction cross section as much as expanding neutron distribution dose for the neutron rich isotopes.
In the figure for 800 MeV, the contribution of proton comes to appear comparing to the figures for lower energies.
Another reason is that the expanding proton distribution gives low density because proton number is fixed with 6, 
and the contribution of such proton is also expected to become small.

\begin{figure}
\begin{center}
{\scalebox{0.7}{\includegraphics{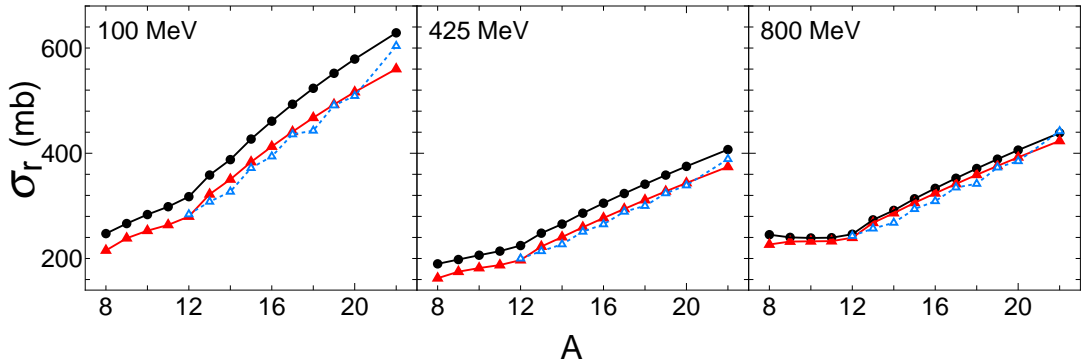}}}
\caption{ Reaction cross sections as the function of the mass number.
Solid circles are results for RIA, and solid triangles  for the Glauber calculation.
Open triangles are corresponding to the results for Ref.\cite{ibr08}.}            
\end{center}
\end{figure}

Figure 4 shows the reaction cross sections with respect to the energy. 
For the RIA, plotted energies are values given in Tab.\ref{t2}.
In  Fig.4, 
the Glauber calculations give significantly large reaction cross section in energies less than 100 MeV, while the RIA calculations do not show significantly increase with decreasing 
energy.
Experimental data have been given in Ref.\cite{kt10} for $^{19,20,22}$C at 40 MeV.
Since the value of incident energy is so small that the RIA calculations are not available, 
the Glauber calculations with the RMF density distributions
are compared with experimental data instead.
The reaction cross section for $^{22}$C at 40 MeV is 810.4 mb in Glauber calculations
while a experimental value is 1338(274) mb which is more than1.5 times larger. 
On the other hand for $^{19,20}$C the reaction cross sections are 732.6 mb 
and 761.1 mb, respectively, and those values are similar to the experimental values of 754(22) mb and 791(34) mb.
As seen in Fig.2 the neutron density of $^{22}$C in Ref.\cite{ibr08} has much larger
root-mean-square radius, and the reaction cross section at 40 MeV is 957 mb,
which is also larger than that of Gulauber calculations with RMF densities.
The experimental data for $^{22}$C seem to suggest that the neutron distribution
of $^{22}$C is much more spreading than the distribution considered in the
present calculations, or at least has much larger root-mean-square radius.

\begin{figure}
\centerline{
{\scalebox{0.55}{\includegraphics{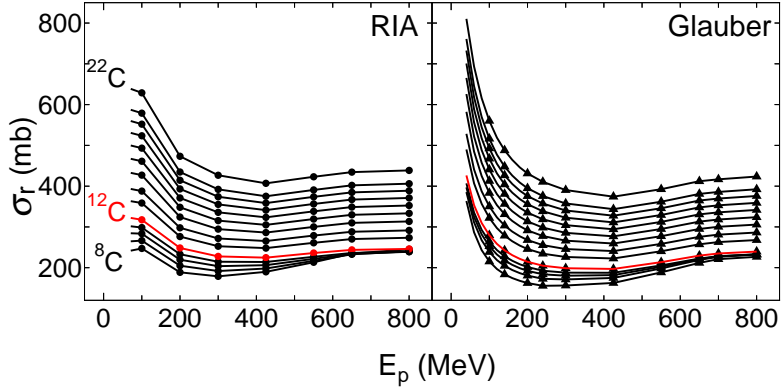}}}}
\caption{ Reaction cross sections with respect to the energy.
The right panel is for RIA, and the left one for the Glauber calculations.}            
\end{figure}

Comparison between results for calculations and the experimental 
data is shown in Fig.5 in the case of $^{12}$C target.
The horizontal axis of energy is logarithmic scale.
The solid line is the result for RIA calculations with tensor density, the dashed line for RIA without tensor density, and dot-dashed line for Glauber calculations.
The solid circles are experimental data taken from Ref.\cite{carl96}.
The Glauber calculations predict the energy dependence of the reaction cross section
overall except for two values at 61 and 77 MeV.
These experimental values seem to be inconsistent with the other data.
The RIA results show good agreement with high energy data, accidentally with low energy
ones. In general RIA calculations give significantly good predictions for proton-elastic
scattering in the energy region higher than 300 MeV.
These results are shown in the following section.

\begin{figure}
\centerline{
{\scalebox{0.45}{\includegraphics{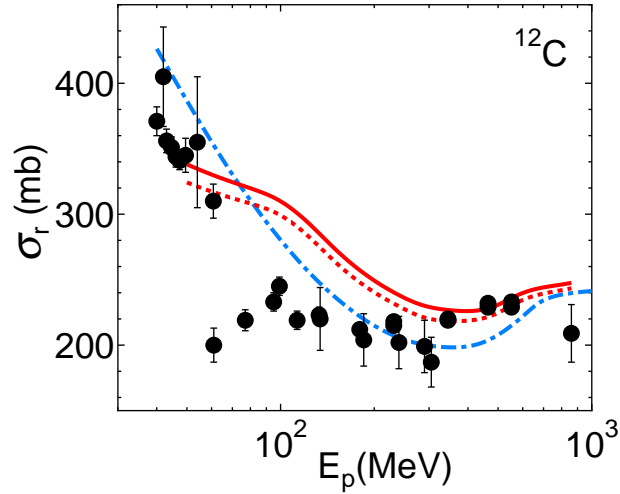}}}}
\caption{ Comparison between results for calculations and the experimental 
data in the case of $^{12}$C target.
The solid line is the result for RIA calculations with tensor density, the dashed line for RIA without tensor density, and dot-dashed line for Glauber calculations.
The solid circles are experimental data taken from Ref.\cite{carl96}.
}
\end{figure}

\subsection{Elastic scattering calculations for RIA}

According to Eq.(20), the ovservables for proton-elastic scattering from carbon isotopes
are calculated with optical potentials based on the RIA, and are compared with 
experimental data.
Figure 6 shows results for proton-$^{12}$C scattering at 150, 250, and 300 MeV;
differential cross section and analyzing power .
The solid line is the result for RIA calculations with tensor density, the dashed line for RIA without tensor density.
Solid circles are experimental data from Ref.\cite{12c150}  (150 MeV), Ref.\cite{12c250}
(250 MeV) and Ref.\cite{12c300}  (300 MeV),  respectively.         
Differential cross sections in the forward angle region: $\theta_{\rm c.m.} \leq$ 40 degrees
are well predicted for all energies shown here.
For such low energy region, it is known that the RIA
predictions for analyzing powers are not so good as
those for differential cross sections, and calculations come to show good agreement
with experimental data in the energies larger than 300 MeV, though such a comparison
is not given in the figure due to absence of the analyzing powre data.
Contributions of tensor densities are small for both differential cross sections and 
analyzing powers in these energies.
\begin{figure*}
\centerline{
{\scalebox{0.60}{\includegraphics{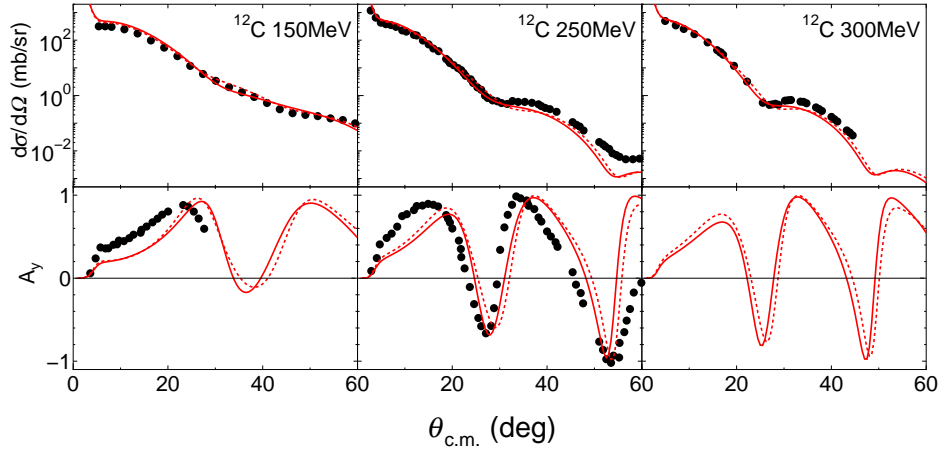}}}}
\caption{ Differential cross section and analyzing power for proton-$^{12}$C
elastic scattering at 150, 250 and 300 MeV.
The solid line is the result for RIA calculations with tensor density, the dashed line for RIA without tensor density.
Solid circles are experimental data from Ref.\cite{12c150}  (150 MeV), Ref.\cite{12c250}
(250 MeV) and Ref.\cite{12c300} (300 MeV),  respectively.}          
\end{figure*}

Figures 7 and 8 show the results for proton-elastic scattering from 
$^{12}$C and $^{13}$C targets.
In Fig.7 ,(a) and (b) are results for 200 MeV, and 800 MeV, respectively.
In each part ,the left side sheets are the results for $^{12}$ C target, 
and the right side ones for $^{13}$C.
The upper sheets show the differential cross section, and the lower sheets the analyzing
power.
The line identifications are the same in Fig.6, and
solid circles are experimental data from Ref.\cite{12c200}  (a), and Ref.\cite{12c800} (b),  respectively.
As already seen in Fig.6, 
the differential cross sections are well predicted in the forward
angle region: $\theta_{\rm c.m.} \leq 40$ degrees at 200 MeV, 
and $\theta_{\rm c.m.} \leq 20$ degrees at 800 MeV.
The analyzing powers at 200 MeV, the angular distribution is similar to the result for
250 Mev, though the calculation of $^{13}$C target shows good agreement with
experimental data in the angle region: 20 $\leq \theta_{\rm c.m.} \leq$ 60 degrees.
The contributions of tensor density is also small at 200 MeV, however, at 800 MeV
they appear in results for the forward analyzing power, and shift the distributions
to larger angles.

\begin{figure*}
\begin{center}
\begin{tabular}[b]{c}
\subfigure[Results for 200 MeV.]
{\scalebox{0.65}{\includegraphics{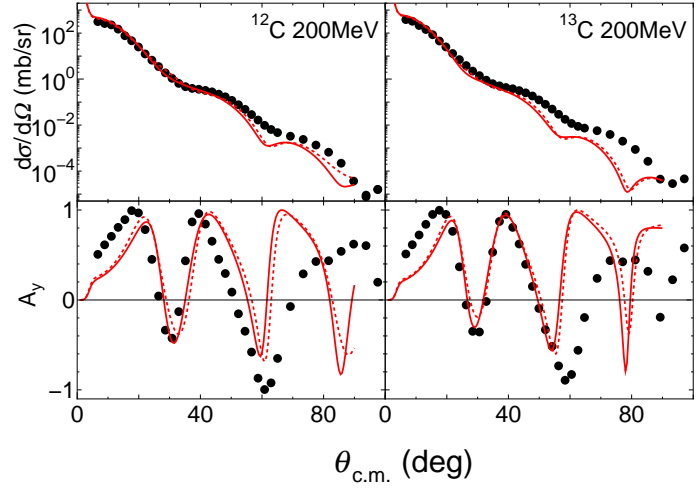}}}  \\

\subfigure[Results for 800 MeV.]
{\scalebox{0.65}{\includegraphics{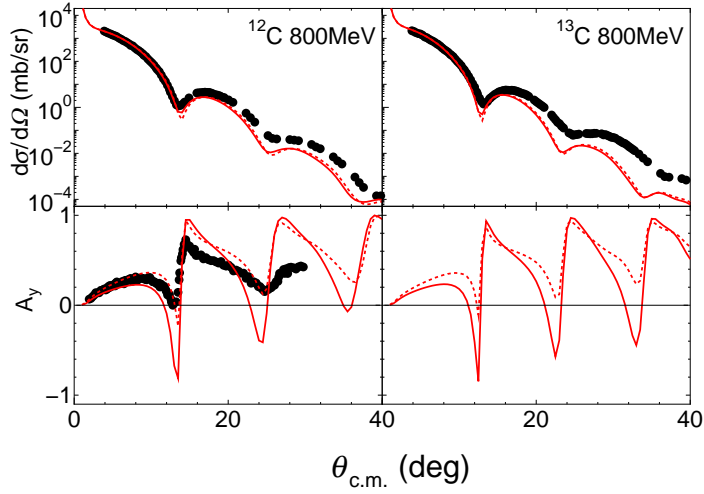}}}
\end{tabular}
\end{center}
\caption{ Differential cross section and analyzing power for $^{12}$C and
$^{13}$C target at 200 MeV (a), and 800 MeV (b).
The left side sheets are results for $^{12}$ C target, and the right side ones for $^{13}$C.
The upper sheets show the differential cross section, and the lower sheets the analyzing
power.
The line identifications are the same in Fig.6, and
solid circles are experimental data from Ref.\cite{12c200}  (a), and Ref.\cite{12c800} (b),  respectively.}            
\end{figure*}

Figure 8 shows differential cross section (a), analyzing power (b)
and spin rotation (c) for $^{12}$C and $^{13}$C target at 500 MeV.
Experimental data given by the solid circles are from Ref.\cite{12c500}, and
the spin rotation shown by R is given $D_{ss}$ in the reference.
The calculated results for differential cross sections predict well in the very forward
region: $\theta_{\rm c.m.} \leq$ 20 degrees, and results for spin obervables 
show good agreement with the data overall, especially the analyzing power for 
$^{13}$C target, which is also seen in Fig.7 (a).
In this case, the contributions of tensor density is significant around the first dip of 
analyzing powers, and make predictions fit to the experimental data.
As seen in Fig. 1 (a), the density distribution of $^{13}$C spreads much more
than that of $^{12}$C though only one neutron exceeds.
The relativistic mean field results show, as given in Tab.I, that
the charge radius of $^{12}$C is smaller than the value of Ref.\cite{am13},
while the charge radii for $^{13}$C and $^{14}$C are slightly lager than 
those of the reference.
In the RIA calculations, the different results between $^{12}$C and $^{13}$C 
originate in the density distributions.
Provided the spreading density distribution of $^{13}$C shows good prediction 
for analyzing power, the elastic scattering data for $^{12}$C are given by 
slightly spreading density of the target nucleus.
In other words, the relativistic mean field result for $^{12}$C in the present calculations
gives rather compact density distribution and may be modified to provide
slightly spreading distribution in order to fit the experimental data.

\begin{figure}
\centerline{
{\scalebox{0.45}{\includegraphics{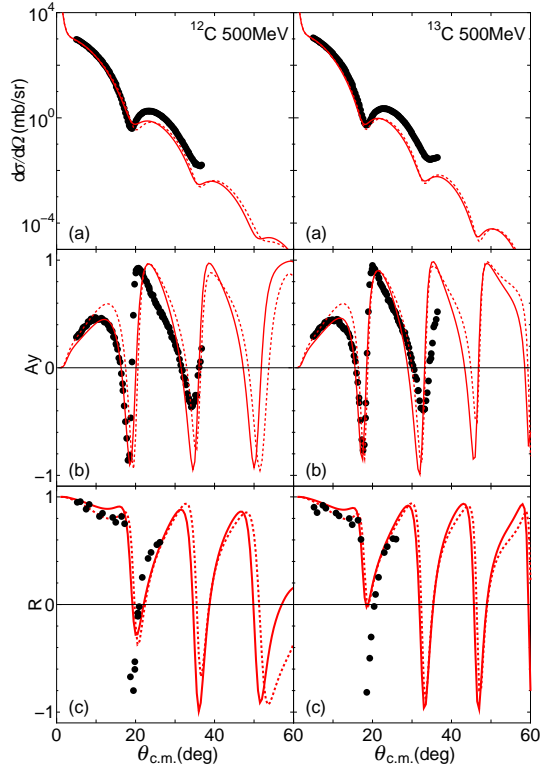}}}}
\caption{ Differential cross section (a), analyzing power (b) 
and spin rotation (c)
for $^{12}$C and
$^{13}$C target at 500 MeV.
The left side sheets are results for $^{12}$ C target, and the right side ones for $^{13}$C.
The upper sheets show the differential cross section, and the lower sheets the analyzing
power.
The line identifications are the same in Fig.6, and
solid circles are experimental data from Ref.\cite{12c500} (494 MeV) .}    
\end{figure}

In Fig.9 appears the differential cross section for  $^{14}$C target at 50 MeV.
The solid circles are experimental data from Ref.\cite{14c40}, while the data
have been taken at 40 MeV.
The proton incident energy 50 MeV is the lowest one for RIA calculation here,
therefore the prediction gives always small values comparing to the experimental data,
even in the forward angle region.
It is however seen that the angular distribution is overall predicted for $^{14}$C target.

In comparison with the experimental data of Ref.\cite{ym13},
though they are not shown here,
the numerical results for $^9$C target at 300 MeV also give small values
as seen in Fig.7.
The root-mean-square radius for nuclear matter in the reference, which has been determined from
the data,  has been 2.43$\pm$0.55 fm, and
this is smaller than the value for the relativistic mean field results given in Tab.I: 2.664
fm while the value itself exists within a margin of error.
As for the differential cross section calculated with RIA, the first dip position seems
to exist in smaller angle than the experimental data.
This phenomenon is consistent with the spreading density distribution of the target
nucleus, e.i., the large root-mean-square radius of the nuclear matter.
In other words, the experimental data seem to prefer the density distribution for $^9$C
with smaller matter radius than the relativistic mean field results.
In the case of small number of neutron, nuclear densities provided by the relativistic
mean field results show a tendency to expand, as seen in helium isotopes \cite{kk14}.

\begin{figure}
\centerline{
{\scalebox{0.45}{\includegraphics{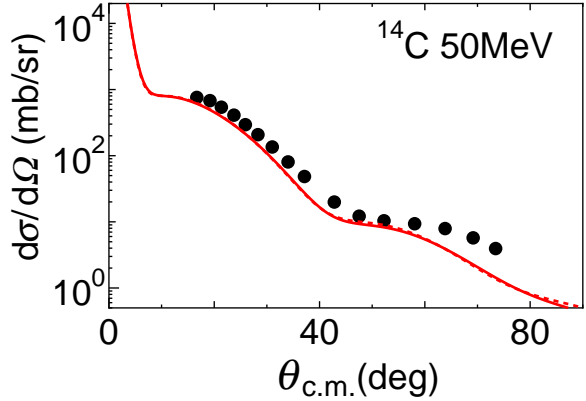}}}}
\caption{ Differential cross section 
for $^{14}$C  target at 50 MeV.
The line identifications are the same in Fig.6, and
solid circles are experimental data from Ref.\cite{14c40} (40 MeV) .}    
\end{figure}

\subsection{Relationship between $r_{rms}$ and $\sigma_r$}

In order to show the relationship between the reaction cross section and
the root-mean-square radius, the density distributions  for $^{12}$C target nucleus
are assumed by Wood-Saxon function as follows;
\begin{eqnarray}
\rho(r) = \frac{\rho_0}{1+\exp[(r-R)/a]},
\end{eqnarray}
where $R$ and $a$ are the half-density radius and diffuseness parameter, respectively.
The value $\rho_0$ is the normalization constant which is determined f
rom the following equation;
\begin{eqnarray}
\int d^3 r \ \rho(r) = N,
\end{eqnarray}
where $N$ is atomic number $Z$ for the proton,  and $A-Z$ for the neutron.
The number $A$ is the mass number of the target nucleus, and for $^{12}$C 
these numbers are the same given by $Z=A-Z=6$.
As usual the half-density radius is given by $R=cA^{1/3}$, therefore $a$ and $c$
are determined freely.
In the present calculations, these parameters are chosen 
so that the root-mean-square radius is the same as the result for the relativistic 
mean filed calculations given in Tab.I, i.e. $\sqrt{<r^2>_p}=2.277$ fm for the proton,  
$\sqrt{<r^2>_n}=2.257$ fm for the neutron, and in result $\sqrt{<r^2>_m}=2.267$ fm
for the nuclear matter, while the deviations of $\pm 0.01$ fm are practically concerned.
The obtained parameter sets are three for proton and neutron, respectively, 
and combinations are nine, which are given in Table 3.
For the diffuseness parameter, $a=0.35, 0.45, 0.55$ fm are first taken, 
and the half-density radii are determined so that the root-mean-square radius is obtained with the value of the relativistic mean-field results.
The distributions of $a=0.45$ fm are similar to the results for the relativistic-mean-field
calculations, therefore the model WS1 corresponds to $^{12}$C in Fig.1 (a).
The results for $a=0.35$ fm are compressed distributions and for $a=0.55$ fm are
spreading ones.

\begin{table}
\caption{Parameter sets for Wood-Saxon function.
}
\begin{center}
\begin{tabular}{ccccc} \hline
model & \multicolumn{2}{c}{proton (fm)}  & \multicolumn{2}{c}{neutron (fm)} \\ \cline{2-5}
        & a     & c     & a  & c   \\ \hline
WS1 & 0.45 & 1.09 & 0.45 & 1.07 \\
WS2 & 0.35 & 1.32 & 0.45 & 1.07 \\
WS3 & 0.55 & 0.73 & 0.45 & 1.07 \\
WS4 & 0.45 & 1.09 & 0.35 & 1.31 \\
WS5 & 0.45 & 1.09 & 0.55 & 0.70 \\
WS6 & 0.35 & 1.32 & 0.55 & 0.70 \\
WS7 & 0.55 & 0.73 & 0.35 & 1.31 \\
WS8 & 0.35 & 1.32 & 0.35 & 1.31 \\
WS9 & 0.55 & 0.73 & 0.55 &0.70 \\ \hline
\end{tabular}
\end{center}
\label{t1}
\end{table}

Figure 10 shows density distributions for the models; WS1 through WS9 
corresponding to Tab.IV.
Solid, dashed, and dash-dotted lines are results for neutron, proton, and 
nuclear matter, respectively.
There is much variety found between density distributions, while they have
the same root-mean-square radii within the error of $\pm 0.01$ fm.

\begin{figure*}
{\scalebox{1.0}{\includegraphics{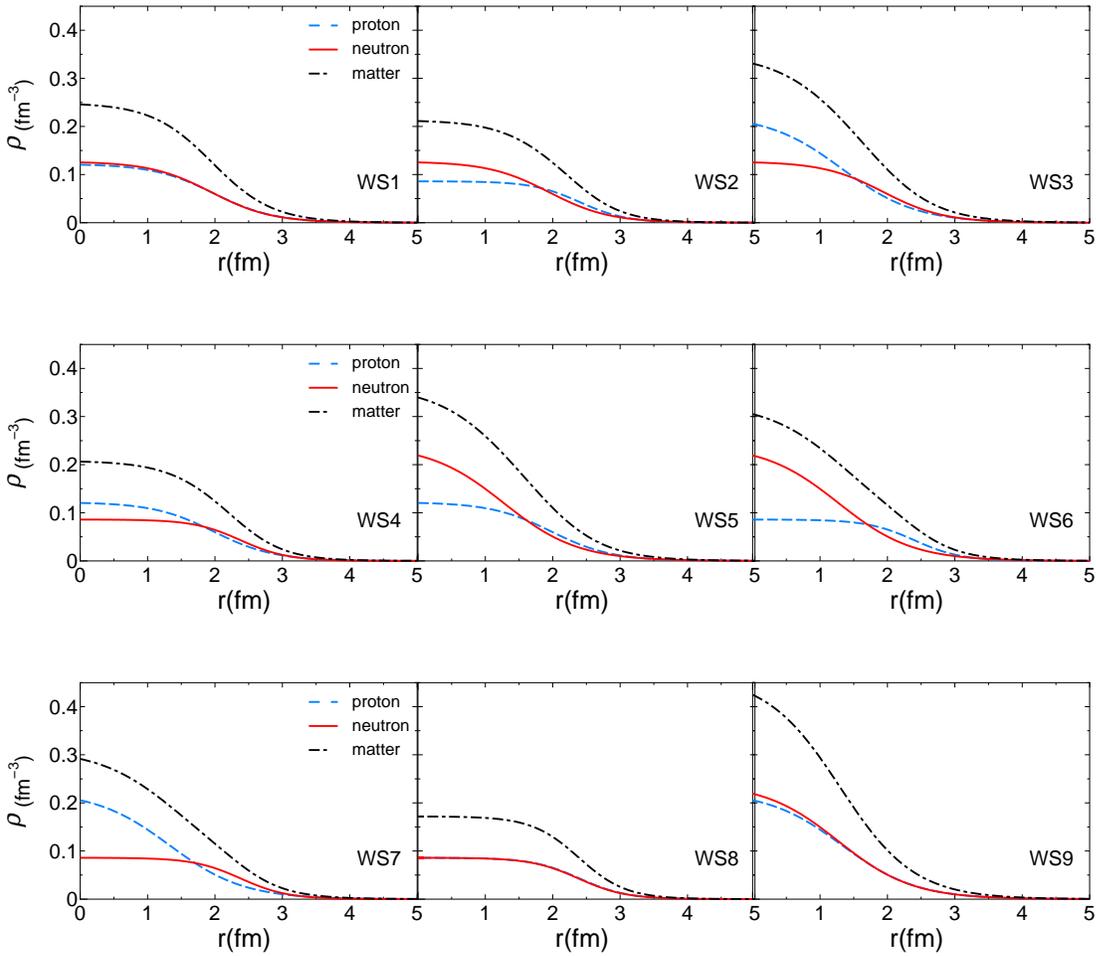}}}
\caption{ Model density distributions.
Solid, dashed, and dash-dotted lines are densities for neutron, proton, and nuclear matter,
respectively.
}          
\end{figure*}

\begin{figure}
\centerline{
{\scalebox{0.60}{\includegraphics{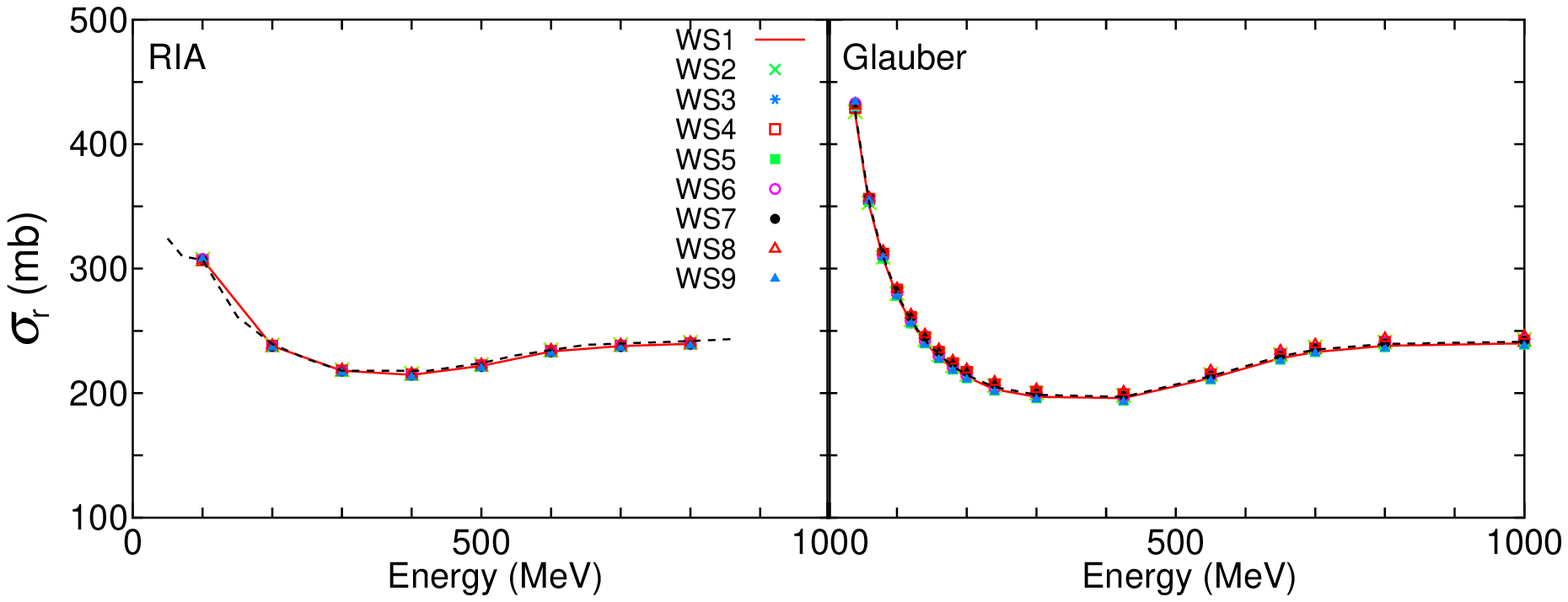}}}}
\caption{ Reaction cross sections calculated with the model density distributions.
The left panel shows the RIA calculations and the right one dose the Glaubar calculations.
In both panels dashed lines correspond to the results for the density distribution 
provided with RMF results.}    

\end{figure}

In terms of these model densities, the reaction cross sections for proton-elastic scattering
from $^{12}$C target are  shown in Fig. 11 in the RIA (left panel) and Glauber (right panel) 
calculations.
In both panels dashed lines correspond to the results for the density distribution 
provided with RMF results.
For the RIA calculations the tensor densities are excluded, and scalar densities
are given by multiplying the model distributions and scalar-vector density ratios
obtained from RMF results.
It is clearly seen that all results for the model WS1 through WS9 provide 
almost the same values in wide energy region calculated here.
As already mentioned, the result for WS1 is expected to be the similar value as
the relativistic-mean field results shown in  Fig. 4.
It is concluded from Fig.11 that there is a strong relationship between
the reaction cross section and the root-mean-square radius, and is also
that the reaction cross sections determine the parameter sets of Woods-Saxon
density distributions, which provide a specific root-mean-square radius.
For calcium and nickel isotopes, 
the reaction cross section and the mean-square
radius have shown almost the same behaviors as 
the functions of parameters for 
Woods-Saxon density distributions based on  the RIA calculations\cite{kk09, kk15}.
The mean-square raius is analytically given by Woods-Saxon density distribution
of Eq. (25) as follows;
\begin{eqnarray}
<r^2> \  = \ \frac{7}{5} ( \pi a )^2 + \frac{3}{5} R^2.
\end{eqnarray}
Figure12 shows the relationship between the half-density radius and diffuseness,
which are given in Tab. IV.
Circles and triangles are results for neutron and proton, respectively.
The solid line corresponds to a part of the ellipse: $8.5=\frac{7}{3}(\pi a)^2 + R^2$.
The number of the left hand side is determined in accordance with the values
which are caluclated with the mean-square radii of proton and neutron as follows; $ (2.277)^2 \times 5/3 \sim 8.64$ and
$(2.257)^2 \times 5/3 \sim 8.49$.
As already mentioned, the half-density radius is searched with respect to the given
diffuseness parameter so that the root-mean-square radius is the same 
as the result for the relativistic mean filed calculations.
Figure 12 confirms that these searched parameter sets completely satisfy Eq.(27).
In order to determine the whole distribution of the target nucleus, another observable
has been considered in the previous works \cite{kk09,kk15}, e.g. the first dip position
of the differential cross section.
Two observables: the reaction cross section and the first dip position of the differential
cross section, in principle have been able to determine two parameters of Woods-Saxon
function while the experimental errors have significantly affected the accuracy in
the determination of the parameters.

\begin{figure}
\centerline{
{\scalebox{0.70}{\includegraphics{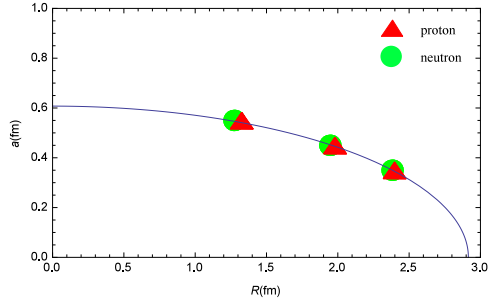}}}}
\caption{ Relationship between the half-density radius and diffuseness.
Circles and triangles are results for neutron and proton, respectively.
The solid line corresponds to a part of the ellipse derived from Eq.(27).}    
\end{figure}

\section{Summary and Conclusion}

This work has presented reaction cross sections for proton-elastic scattering
from carbon isotopes of $A=8-22$ except $A=21$  in large energy region: 100-800 MeV .
Density distributions of the target nuclei have been provided from the relativistic
mean-field results, and calculations have been done in terms of 
the relativistic impulse approximation.
As for reference, Glauber model calculations with RMF density distributions
have been also given.

Reaction cross sections which have been calculated with RIA are sightly larger
than those with the Glauber calculations in the whole energy region considered here.
The behavior with respect to the energy is similar in both calculations, i.e.
significantly decreasing with increasing energies smaller than 200 MeV,  
showing minimum values at around 300-400 MeV, 
and after that slightly increasing with increasing energies.
These phenomena are mainly attributed to the NN amplitudes on which both
prescriptions have based in the calculations.

As expected, the reaction cross sections increase with increasing mass
number of carbon isotopes, however, the root-mean-square radius
shows much larger value for the isotopes whose mass numbers are less than 12
due to the expanding proton distributions.
Such expansion is caused by both repulsive Coulomb interaction
and small number of neutron which gives rise to attractive nuclear interaction.
Contributions of expanding proton distributions have been slightly seen
while those of neutron distributions have significantly appeared.
For the proton-rich isotopes,
effects of decreasing mass number and increasing root-mean-square radius
contribute to the reaction cross section in the opposite direction each other.
Therefore it is rather complicated to find the direct relationship between 
$\sigma_r$ and $r_{ \rm rms}$.
In the case of the neutron-rich isotope, the root-mean-square radius
simply increases with increasing mass number, and the relation of $\sigma_r$
to $r_{ \rm rms}$ is expected to be a plain one.

In order to show the relationship between $\sigma_r$ and $r_{ \rm rms}$,
a model analysis with Woods-Saxon density distributions for $^{12}$C nucleus
has been done.
It has been shown that various distributions with different parameters
provided almost the same values of the reaction cross in the large
energy region: 100-800 MeV as far as the distributions had the
same values of the root-mean-square radius.
Such a strong relationship between $\sigma_r$ and $r_{ \rm rms}$ 
provides some prescriptions which determine the root-mean-square radius
directly from the reaction cross section at least for the neutron rich nuclei.
Besides the reaction cross section, however, 
another observable is necessary  to obtain the whole profile of the density distribution
for the target nucleus in the proton-elastic scattering.
For another possibility, reaction cross sections in nucleus-nucleus scattering 
are expected to determine the density distributions 
though the RIA calculations become rather complicated but are challenging.

\section{acknowledgments}
The author acknowledges the use of the code of the relativistic
mean field calculation provided from Y. Sugahara.  
Numerical calculations in this paper were performed using the facilities at the Information Processing Center of Shizuoka University, and partly using the computing service at
Institute for Information Management and Communication, Kyoto University.
Some of numerical calculations in Sect. 3.4 have partly appeared in Ref. \cite{km14}


\end{document}